# Manuscript Template

## FRONT MATTER

### Title
- [Full] Programmable and sequential Gaussian gates in a loop-based single-mode photonic quantum processor
- [Short] Gaussian gates in a loop optical quantum processor

### Authors
Yutaro Enomoto,[1] Kazuma Yonezu,[1] Yosuke Mitsuhashi,[1] Kan Takase,[1] Shuntaro Takeda[1,2]*

### Affiliations
[1] Department of Applied Physics, School of Engineering, The University of Tokyo, 7-3-1 Hongo, Bunkyo-ku, Tokyo 113-8656, Japan.
[2] JST, PRESTO, 4-1-8 Honcho, Kawaguchi, Saitama 332-0012, Japan.
*Corresponding author. Email: takeda@ap.t.u-tokyo.ac.jp

### Abstract
A quantum processor to import, process, and export optical quantum states is a common core technology enabling various photonic quantum information processing. However, there has been no photonic processor which is simultaneously universal, scalable, and programmable. Here, we report on an original loop-based single-mode versatile photonic quantum processor which is designed to be universal, scalable, and programmable. Our processor can perform arbitrarily many steps of programmable quantum operations on a given single-mode optical quantum state by time-domain processing in a dynamically controlled loop-based optical circuit. We use this processor to demonstrate programmable single-mode Gaussian gates and multi-step squeezing gates. In addition, we prove that the processor can perform universal quantum operations by injecting appropriate ancillary states and also be straightforwardly extended to a multi-mode processor. These results show that our processor is programmable, scalable, and potentially universal, leading to be suitable for general-purpose applications.

### Teaser
A loop-based photonic quantum processor for general-purpose applications is demonstrated to be scalable and programmable.

## MAIN TEXT

### Introduction

A quantum processor capable of performing universal quantum operations on a given arbitrary optical quantum state is a common core technology enabling various photonic quantum information processing. Requirements for such a general-purpose quantum processor are (i) the basic functionality to import, process, and export optical quantum states, (ii) ability to perform a universal set of operations, (iii) scalability to efficiently perform arbitrary steps of operations, and (iv) programmability to perform various tasks. Such a processor is at the heart of continuous-variable quantum computing (*1, 2*) as well as pre-/post-processing in quantum communication and sensing protocols (*3–6*). It is also an essential building block to realize Hamiltonian simulation for various quantum systems (*7–9*) and quantum neural networks for machine learning applications (*10–12*).

Thus far, most of the elementary building blocks for photonic quantum processors have been demonstrated separately. In general, universal quantum processors are required to perform arbitrary Gaussian operations (such as squeezing and phase shift gates) for linear transformation and at least one non-Gaussian operation (such as a cubic phase gate) for nonlinear transformation. Sequential application of Gaussian and non-Gaussian operations is proven to constitute arbitrary transformations (*2*). Gaussian operations alone are already useful for many quantum communication and sensing protocols (*13, 5, 6*), while non-Gaussian operations are proven to be necessary for quantum computational speedup over classical computers (*14*). On the experimental side, optical Gaussian and non-Gaussian operations are efficiently implementable with "measurement and feedforward" schemes (*15–17*). Proof-of-principle demonstrations of such single-step Gaussian gates have been reported already (*18–20*). Recent research interests are moving on to integrating these building blocks to realize general-purpose quantum processors.

However, to date, no photonic processor has been demonstrated that simultaneously satisfies all the requirements (i)-(iv) for the general-purpose processor. In the context of continuous-variable quantum computing, multi-step quantum operations have been demonstrated with the "measurement and feedforward" schemes (*21, 22*), but these early demonstrations relied on spatial-mode encoding, making them difficult to scale up. These schemes were later extended to temporal-mode encoding to achieve high scalability (*23–25*), and programmable multi-step quantum computations have been recently reported with temporal mode cluster states (*26, 27*). However, these implementations were designed specifically for the purpose of one-way quantum computing, which internally prepares input quantum states for computation and returns only classical calculation results instead of output quantum sates. Thus, these implementations lack essential ingredients for general-purpose quantum processors such as an input port to take in external states or optical feedforward to finalize quantum operations.

Here, we report on an original loop-based single-mode versatile photonic quantum processor which is designed to satisfy all the requirements for the general-purpose processor. Our loop-based processor can store an externally injected single-mode optical state (*28, 29*) and perform arbitrarily many steps of quantum operations by repeating a "measurement and feedforward" process in the loop. This time-domain processing is achieved by developing dynamical and synchronized control of the circuit parameters at a nanosecond time-scale, such as transmissivity of beam splitters, measurement bases, and availability of feedforward. The dynamical controllability also allows us to fully program gate sequences without changing the hardware configurations. In this work, we use this processor to demonstrate programmable single-mode Gaussian gates and multi-step squeezing gates. Moreover, we prove that the processor can perform a non-Gaussian gate by injecting appropriate ancillary states. The non-Gaussian gate is not implemented here, but its in-principle applicability and the multi-step operationability demonstrated in this work show that our processor can perform universal single-mode quantum operations by sequential Gaussian and non-Gaussian gates. These results together show that our processor is scalable, programmable, and potentially universal, leading to be suitable for general-purpose applications. This work can be straightforwardly extended to a multi-mode processor (*28*), and will accelerate the realization of various quantum optical applications as a universal processing unit.

## Results
### Working principle of the loop-shaped processor
The basic action of our processor is to let the input wave packet circulate in a loop and repeatedly perform measurement-induced Gaussian/non-Gaussian gates with the help of ancillary states (*28*). Figure 1A shows the conceptual diagram of the loop-shaped optical circuit. The input state to the circuit is in the form of a single wave packet. The wave packet of each ancillary state is sequentially injected into the circuit via an optical switch (Switch-1) that transmits either the input or the ancillary states. These states are then coupled to the loop by a variable beam splitter (VBS), the transmissivity of which is dynamically controlled. In the loop part, displacement and phase shift operations are performed. The amount of displacement and phase shift operation is also dynamically controlled, which is represented by time dependent parameters of $g(t)$ and $\theta(t)$ (Fig. 1A). At the downstream of the loop part, there is another optical switch (Switch-2). One output connection of Switch-2 leads to a homodyne detector with a dynamically tunable measurement basis; $\hat{x}\cos\phi(t) + \hat{p}\cos\phi(t)$ is measured, where $\hat{x}$ and $\hat{p}$ are the quadrature operators of the light field and $\phi$ is called the homodyne angle. The measurement outcome of the homodyne detector is fed forward to the displacement operation inside the loop when necessary. The other connection of Switch-2 is coupled to free space, which serves as the output port of the whole processor.

Thanks to the dynamical feature of each component, the loop-shaped circuit can be virtually transformed into a sequential optical circuit, which is depicted in Fig. 1B. This figure essentially shows that single time-dependent components can be viewed as multiple time-independent components. In the case of the VBS, for example, the single VBS is interpreted as the multiple beam splitters (BSs) having independent constant values of reflectivity, which are named BS-0 – BS-($n$ + 1). As can be seen from Fig. 1B, the processor is capable of applying measurement-induced operations arbitrarily many times to the input state. Moreover, such sequential operations are sufficient for universal single-mode operations. In fact, except for already built-in displacement operations, the processor in Fig. 1A can implement any single-mode Gaussian gate by sequential operations of a phase shift, measurement-based squeezing with an ancillary squeezed state, and another phase shift (*30*). A cubic phase gate, as an example of non-Gaussian gates, can also be performed by using proper ancillae based on a scheme we propose in this work (refer to Materials and Methods section). This scheme is an alternative to the one in the reference (*28*) and has the advantage that a cubic phase gate is performed with the single loop optical path instead of the dual loop. Therefore, the processor can perform universal single-mode operations by sequential Gaussian/non-Gaussian gates (*2*) with appropriate ancillary states. In addition to the above property, the operations by the processor are programmable as well since the parameters of the circuit are dynamically controlled from outside.

### Single-step Gaussian gate
In order to demonstrate the programmability of the processor, we perform several different single-step Gaussian gates by the measurement-based scheme in reference (*16*), which first couples an input state to an ancillary squeezed state, then measures one part of the state, and finally performs feedforward to the other part. As representatives among all the Gaussian gates, we choose squeezing gates and quadratic phase gates (QPGs). Input-output relations of these gates can be expressed in a matrix form as $[\hat{x}_{\text{out}}, \hat{p}_{\text{out}}]^T = S(r)[\hat{x}_{\text{in}}, \hat{p}_{\text{in}}]^T$ or $[\hat{x}_{\text{out}}, \hat{p}_{\text{out}}]^T = U_2(\kappa)[\hat{x}_{\text{in}}, \hat{p}_{\text{in}}]^T$ with

$$S(r) = \begin{bmatrix} e^r & 0 \\ 0 & e^{-r} \end{bmatrix} \quad \text{(squeezing gate)} \tag{1}$$

$$U_2(\kappa) = \begin{bmatrix} 1 & 0 \\ 2\kappa & 1 \end{bmatrix} \quad \text{(QPG)} \tag{2}$$

where $\hat{x}_{\text{in}}$ and $\hat{p}_{\text{in}}$ are the quadrature operators (refer to Materials and Methods section for notations and conventions) of the input state while $\hat{x}_{\text{out}}$ and $\hat{p}_{\text{out}}$ are those of the output state. Let us call $r$ and $\kappa$ as a squeezing parameter and a QPG parameter, respectively. We choose a squeezing parameter as $r = 0.44$ and $0.69$, and a QPG parameter as $\kappa = 0.46$ and $0.75$ for the demonstration.

The operation condition of the system can be described by using Fig. 1B; the roundtrip number $n$ is one for the single-step gate. We inject a vacuum field into the system as an input resource, and then, by utilizing the displacement and the phase operations in the loop, we prepare three different types of input coherent states $|\alpha\rangle$ with $\alpha = 0$, $x_{\text{in}}$, and $ip_{\text{in}}$ ($x_{\text{in}}, p_{\text{in}} \in \mathbb{R}$). Let us call these input states as the vacuum state, X-coherent state, and P-coherent state for later convenience. Ancilla-1 is the squeezed vacuum state. The amount of phase shifts $\theta_1$ and $\theta_2$ and the transmissivity of the VBS are dynamically set to appropriate values according to the desired squeezing parameter $r$ or QPG parameter $\kappa$ (Table I). In all cases, the homodyne angle $\phi_1$ is 0 deg. After the input state is processed by the squeezer and the phase shifters, the reflectivity of the BS is changed to zero (BS-($n$+1) in Fig. 1B) so that the wave packet comes out of the loop.

Figure 2A describes the experimental setup for the demonstration. Each component is dynamically controlled, and the timing of the control is synchronized at a nanosecond time-scale by using a timing controller. Figure 2B shows how such a dynamical and synchronized control is typically achieved in the demonstration. What makes it possible for the optical circuit to perform quantum logic gates in the time domain is to incorporate dynamical and synchronized control of the input switch (Switch-1) and feedforward system into our previous work (*29*), which is the major technical advance in this work. In our setup shown in Fig. 2A, with Switch-2 omitted, the output of the processor is measured by the homodyne detector, where the output state is characterized by a tomography process with the local oscillator phase varied (*31*). The detail of the setup is described in Materials and Methods section.

Figure 3 summarizes the results of operation of the squeezing gates with $r = 0.44$ and $0.69$. Figure 3A shows the output Wigner functions of the squeezing gates performed on the X-coherent input state, as an example out of three different input states. The Wigner function of the input state is also shown. According to Eq. 1, the gate operation changes the standard deviations along $x$- and $p$-axes by factors of $e^r$ and $e^{-r}$, respectively. In the same way, the mean values along $x$- and $p$-axes are also changed by the same factors of $e^r$ and $e^{-r}$, respectively. Thus, the shapes of the output Wigner functions are squeezed in the direction of the $p$-axis while the positions are shifted along the $x$-axis, following the matrix transformation of Eq. 1. It can be seen from these plots that the positions and shapes of the ellipses of the experimental outputs agree well with those of the theoretical ones (Fig. 3B), which indicates that the processor performs the squeezing gates as expected. The experimental and theoretical outputs agree well also for the vacuum and P-coherent input states (Supplementary Materials). Figures 3C and 3D show the comprehensive evaluation of the gates by quantitatively comparing the experimental and theoretical output states. Figure 3C shows matrix elements of the squeezing gates $S(r)$ inferred from transformation of mean values of the X- and P-coherent input states. The (1, 1)- and (2, 1)-elements are estimated by $\langle\hat{x}_{\text{out}}\rangle/\langle\hat{x}_{\text{in}}\rangle$ and $\langle\hat{p}_{\text{out}}\rangle/\langle\hat{x}_{\text{in}}\rangle$, respectively, with the X-coherent input state injected into the processor. Here, $\langle\cdots\rangle$ denotes the mean value. Similarly, the (1, 2)- and (2, 2)-elements are estimated by $\langle\hat{x}_{\text{out}}\rangle/\langle\hat{p}_{\text{in}}\rangle$ and $\langle\hat{p}_{\text{out}}\rangle/\langle\hat{p}_{\text{in}}\rangle$, respectively, with the P-coherent input state. The matrix elements derived from the experiment coincide well with the theoretical expectations. Figure 3D shows variances along major and minor axes of squeezing ellipses of the output states and tilt angles of the ellipses. The vacuum input state is used for the estimation of the variances. These results also agree well with the theoretical expectations. The fidelity $F$ between the experimental and ideal output states is $F = 0.86(2)$ and $0.75(2)$ for $r = 0.44$ and $0.69$, respectively, while the realistic model of our setup including optical loss and finite squeezing predicts $F = 0.89$ and $0.80$, respectively. As a whole, the results shown in Fig. 3 together demonstrate that the squeezing gates are properly performed by the processor.

In the same manner, Fig. 4 summarizes the results of the operation of the QPGs with $\kappa = 0.46$ and $0.75$, showing that the QPGs are also properly performed. Comparing Figs. 4A and 4B tells us that the positions and shapes of the output Wigner functions agree well with what the theoretical model predicts (refer to Supplementary Materials for the vacuum and P-coherent input states). The agreement with the model can also be confirmed by Figs. 4C and 4D. The fidelity takes the value of $0.87(3)$ and $0.77(1)$ for $\kappa = 0.46$ and $0.75$, respectively, while the realistic model predicts $F = 0.89$ and $0.80$, respectively.

The deviations between the experimental results and theoretical expectations found in Figs. 3 and 4 can be attributed to several reasons: suboptimal feedforward gain, systematic errors in the phase shift operations, and optical losses that are not taken into account in the theoretical model.

It is worth emphasizing here that the same circuit implements all the above gate operations without any changes of the hardware configuration. Therefore, these results demonstrate programmability of the present processor and capability of applying arbitrary Gaussian gates to a single-mode input.

**Multi-step squeezing gate**
In order to show capability of applying gates multiple times, we perform multi-step squeezing gates. For the demonstration, a squeezing gate with $r = 0.44$ is applied $n$ times ($n = 1, 2,$ and $3$) to three different input states of the vacuum, X-coherent, and P-coherent states. The operation condition and the experimental setup are the same as the previous single-step squeezing gate except that settings for the squeezing process are held over multiple roundtrips of the input wave packet. We choose to apply the same squeezing gate over the multiple steps for the demonstration, rather than to apply different gates for each step, due to the limitation of our current EOM driver for EOM-2 (refer to Supplementary Materials for more information).

Figure 5 shows the results of the demonstration of the multi-step squeezing gates. For the X-coherent input state, the Wigner functions of the experimental output of the $n$-step squeezer are shown in Fig. 5A, while the model expectations are in Fig. 5B. The Wigner functions for the vacuum and P-coherent input states are shown in Supplementary Materials. They graphically show that the multi-step gates are properly performed. The transformation of the mean value and variance is characterized as shown in Figs. 5C and 5D, which also show good agreement with the theoretical expectations. The fidelity takes the value of $0.86(2), 0.67(3),$ and $0.48(3)$ for $n = 1, 2,$ and 3, respectively, while the realistic model predicts $F = 0.89, 0.73,$ and $0.54$, respectively. The deviations in the $(1,1)$-element in Fig. 5C can be attributed to the spatial mode mismatch during the multiple roundtrips in the loop part. The results demonstrate that the loop-shaped processor is able to perform multi-step gates on a given single-mode input in a programmable manner. It should be emphasized here that this setup can operate arbitrarily many steps of Gaussian gates in principle, though a squeezing gate is performed up to three times in the demonstration.

**Discussion**

In conclusion, our experiment has demonstrated that the loop-shaped processor is programmable, capable of performing an arbitrary Gaussian gate, and compatible with multi-step gate operations. The processor can perform arbitrarily many steps of quantum gates in principle and thus work as a universal processor for a single-mode input, once a cubic phase state is provided as ancillary non-Gaussian resources.

It must be emphasized here again that our implementation of quantum gates includes the dynamical feedforward system so that the processor imports and exports photonic quantum states; in contrast, the previous implementations (*26, 27*) which were designed specifically for the purpose of one-way quantum computing did not have an input port or a feedforward system and thus cannot be a general-purpose photonic processor that receives, processes, and returns photonic quantum states. In fact, with Switch-2 as depicted in Fig. 1 to send the output state to the free space, the processor could be a core part of the various applications. Moreover, the processor can be extended to a multi-mode input and output straightforwardly by employing an additional loop optical path outside the current processor as shown in the original proposal (*28*). In this way, our processor will be evolved into a general-purpose photonic processor, which implements arbitrary photonic quantum information processing.

**Materials and Methods**
**Experimental setup**
Figure 2A illustrates the experimental setup of the demonstration. The basic working principle and the functionality of a part of the setup are explained in the reference (*29*). The ancillary squeezed vacuum is produced by an optical parametric oscillator (OPO) that is pumped with a continuous-wave laser beam having wavelength of 430 nm. The other beams have wavelength of 860 nm. A set of EOM-1 and polarizing beam splitters (PBSs) forms Switch-1. In a similar way, a set of EOM-2, PBSs, and a quarter-wave plate (QWP) works as a VBS (refer to Supplementary materials for more information). Displacement operation is performed by coupling a phase-modulated coherent field (displacing beam) by a high-reflectivity mirror inside the loop. The carrier field of the displacing beam is undesired and thus removed by introducing a Mach–Zehnder interferometer. EOM-5 on the path of the displacing beam is used for the feedforward operation ((v) in Fig. 2) as well as the input state preparation ((iii) in Fig. 2), in which a vacuum field from outside the processor is displaced to produce a coherent state. The experimental details around the displacing beam are explained further in Supplementary Materials. The phase shift operation is performed by EOM-3. The output of the processor is measured by the homodyne detector with the homodyne angle varied, where EOM-4 shifts the propagation phase of the local oscillator (LO) beam. The 3-dB bandwidth of the homodyne detector is 200 MHz. All the dynamical components are synchronized by using a timing controller. Figure 2B represents how the parameter of each component is dynamically and synchronously changed. The rise/fall time of EOMs for switching is ~ 10 ns. This plot is produced by an auxiliary measurement where classical light fields are introduced into the optical circuit and detected at the output port so that the light fields probe the dynamical changes of the components. This plot corresponds to the case of a QPG with $\kappa = 0.75$ on the X-coherent input state. Since the loop has roundtrip length of 19.8 m, which corresponds to roundtrip time of 66 ns, individual optical modes are defined every 66 ns. Thus, the parameter of each optical and electrical component is changed with unit time being 66 ns, except for (iii) State preparation (Supplementary Materials).

**Data analysis**
For quantum tomography of the output states, the outcome of the homodyne detector is acquired by an oscilloscope, and then stored in a computer. Among a full stretch of a single time bin, 66 ns, the central 46 ns is used for defining a wave packet mode, while the remaining part is regarded as switching time and thus discarded. We obtain a single sample of a quadrature amplitude by applying a mode function $f(t - t_0)$ to the acquired time series. The mode function is defined by

$$f(t) = \begin{cases} t e^{-\gamma^2 t^2} & (|t| < t_1) \\ 0 & \text{(otherwise)} \end{cases} \quad (3)$$

where $\gamma = 6 \times 10^7 /\text{s}$ and $t_1 = 23$ ns. $t_0$ is set to the center of the time bin. The main benefit of using this mode function is its insensitivity to the overall offset (*32*), which is difficult to remove in many experimental situations. The raw value of a quadrature is calibrated by measuring a vacuum field immediately after the main measurement, so that the raw value is converted into a physical value.

For each value of the homodyne angle, we sample 1000 independent quadrature amplitudes by repeating the same operations of the processor. Twelve different values of the homodyne angle are chosen in increments of 15 deg. As a whole, the tomography of the output state is performed using $12 \times 1000$ samples. The Wigner function is estimated by the method of maximum likelihood with the assumption that the Wigner function is Gaussian (*31*). This assumption is justified in our case because a Gaussian operation never transforms a Gaussian state into a non-Gaussian state. By splitting 1000 samples into ten individual subsets and estimating the Wigner function for each subset, we obtain mean values and statistical errors of the parameters of the Wigner functions.

**Optical loss and OPO**
Let us here describe the model parameters for our experimental setup, which are used to produce the theoretical predictions in Figs. 3–5. The quality of the squeezed state from the OPO is modeled in the following way. The amount of optical loss between the OPO and the VBS is measured to be 7 %, including the internal loss inside the OPO. The roundtrip optical loss in the loop part is measured to be 6 %. The readout loss is measured to be 19 %, including the optical loss between the VBS and the homodyne detector as well as imperfection in the homodyne detection such as mode mismatching effect of the LO field and electrical noise. The squeezing level of the pure squeezed state from the OPO, before suffering from any optical loss, is estimated to be 8.0 dB by measuring the squeezed state from the OPO with the loop part bypassed and considering the losses of 7 % and 19 % in the path from the OPO to the homodyne detection. The realistic model including the optical loss is derived based on these model parameters. In the ideal model, each value of the loss is set to zero and the squeezing level of the ancillary state from the OPO is assumed to be infinity.

**Continuous-variable gate conventions**

In this sub-section, let us summarize the conventions concerning the continuous-variable gates. $\hat{a}^\dagger$ and $\hat{a}$ are creation and annihilation operators of an optical mode, respectively, which satisfy $[\hat{a}, \hat{a}^\dagger] = 1$. We use $\hat{x}$ and $\hat{p}$ for corresponding quadrature amplitudes, which are defined by $\hat{x} = \sqrt{\hbar/2}\,(\hat{a} + \hat{a}^\dagger)$ and $\hat{p} = \sqrt{\hbar/2}(-i\hat{a} + i\hat{a}^\dagger)$. A squeezing gate, firstly, is characterized by a unitary operator $\hat{S}(r) = \exp[(r/2)(\hat{a}^{\dagger 2} - \hat{a}^2)]$. In the Heisenberg picture, the input-output relation of $\hat{S}(r)$ can be written in a matrix form as

$$\begin{bmatrix} \hat{x}_{\text{out}} \\ \hat{p}_{\text{out}} \end{bmatrix} = \begin{bmatrix} e^r & 0 \\ 0 & e^{-r} \end{bmatrix} \begin{bmatrix} \hat{x}_{\text{in}} \\ \hat{p}_{\text{in}} \end{bmatrix} \tag{4}$$

where input and output quadratures are labelled by subscriptions of "in" and "out", respectively.

Let us then define a QPG by a unitary operator $\hat{U}_2(\kappa) = \exp[i\kappa \hat{x}^2/\hbar]$. In a similar way, the input-output relation of $\hat{U}_2(\kappa)$ can be written as

$$\begin{bmatrix} \hat{x}_{\text{out}} \\ \hat{p}_{\text{out}} \end{bmatrix} = \begin{bmatrix} 1 & 0 \\ 2\kappa & 1 \end{bmatrix} \begin{bmatrix} \hat{x}_{\text{in}} \\ \hat{p}_{\text{in}} \end{bmatrix} \tag{5}$$

Here, it is worth mentioning that the following matrix identity holds:

$$\begin{bmatrix} 1 & 0 \\ 2\kappa & 1 \end{bmatrix} = \begin{bmatrix} \cos\theta_2 & -\sin\theta_2 \\ \sin\theta_2 & \cos\theta_2 \end{bmatrix} \begin{bmatrix} e^r & 0 \\ 0 & e^{-r} \end{bmatrix} \begin{bmatrix} \cos\theta_1 & -\sin\theta_1 \\ \sin\theta_1 & \cos\theta_1 \end{bmatrix} \tag{6}$$

where $\kappa = \sinh r$, $\cos 2\theta_1 = \tanh r$, $\sin 2\theta_1 = -1/\cosh r$, and $\theta_2 = \theta_1 + \pi/2$. This identity proves that a QPG can be realized by a successive operation of a phase shifter with $\theta_1$, a squeezing gate with $r$, and another phase shifter with $\theta_2$.

Throughout the paper, we adopt $\hbar = 2$ so that the variance of quadrature amplitudes is equal to unity for the vacuum state.

**Cubic phase gate**

We explain the implementation scheme of a cubic phase gate with the loop-based processor here. Let us first define a cubic phase gate by a unitary operator $\hat{U}_3(\kappa) =$ $\exp[i\gamma \hat{x}^3/\hbar]$. The quadrature operators of the output of the gate can be written in terms of those of the input as

$$\hat{x}_{\text{out}} = \hat{x}_{\text{in}} \quad (7)$$
$$\hat{p}_{\text{out}} = \hat{p}_{\text{in}} + 3\gamma \hat{x}_{\text{in}}^2 \quad (8)$$

As originally proposed in the thesis (*33*), a cubic phase gate can be implemented by a measurement-induced method using a cubic phase state and a squeezed state as ancillae. Figure 6 shows the overview of the method. The input wave packet has interaction with the cubic phase state at BS-1. One outgoing beam from BS-1 is measured by HD-1, where the quadrature operator $\hat{x}$ is measured, while the other is transferred to BS-2. Then the squeezed state is transferred to BS-2 as an ancilla. One outgoing beam from BS-2 is measured by HD-2, where the specific linear combination of the quadrature operators, $\hat{p} + 2\mu\hat{x}$, is measured. Let $y$ denote the measurement outcome of HD-2. Here, the parameter $\mu$ is determined by the measurement outcome of HD-1 $q$ as

$$\mu = 3\gamma \frac{R}{T^{3/2}} q \quad (9)$$

where $R$ and $T$ are the reflectivity and transmissivity of the beam splitters, respectively. The other beam from BS-2 becomes the output of the whole gate after the displacement operation is performed on it. The amount of the displacement is as follows:

$$\hat{x} \rightarrow \hat{x} + g_1 q \quad (10)$$
$$\hat{p} \rightarrow \hat{p} + g_2 y + g_3 q^2 \quad (11)$$

where $g_1 = \sqrt{T}$, $g_2 = \sqrt{T/R}$, and $g_3 = 3\gamma(T-R)/T$. As a whole, the output of this circuit can be expressed with respect to the quadratures of input and ancillary states as

$$\hat{x}_{\text{out}} = \hat{x}_{\text{in}} + \sqrt{T}\hat{x}_{\text{sq}} \quad (12)$$

$$\hat{p}_{\text{out}} = (\hat{p}_{\text{in}} + 3\gamma\hat{x}_{\text{in}}^2) + \left[\sqrt{\frac{T}{R}}\hat{p}_{\text{CPG}} - 3\gamma\left(\sqrt{\frac{R}{T}}\hat{x}_{\text{CPG}}\right)^2\right]$$
$$- 6\gamma\left(\frac{R}{\sqrt{T}}\hat{x}_{\text{in}} - \frac{R^{3/2}}{T}\hat{x}_{\text{CPG}}\right)\hat{x}_{\text{sq}} \quad (13)$$

where the subscriptions, "sq" and "CPG", denote the quadratures of the squeezed state and the cubic phase state, respectively. Here, $\hat{x}_{\text{sq}}$ approaches to zero as the squeezing level of the ancillary squeezed state approaches infinity. Under this condition, $\hat{x}_{\text{out}}$ approaches to the ideal form (Eq. 7). Similarly, the second term of $\hat{p}_{\text{out}}$ approaches to zero since it takes the form of the nullifier of a cubic phase state; the reflectivity and transmissivity of the beam splitters are chosen according to the given $\gamma$ and the ancillary cubic phase state. The third term can be made arbitrarily small by carefully managing the quality of the two ancillary states, though the variance of $\hat{x}_{\text{CPG}}$ is infinity for the ideal ancillary cubic phase state. In summary, this measurement-induced gate approaches the cubic phase gate described in Eqs. 7 and 8 in the limit of pure ancillary states. Apparently, the structure shown in Fig. 6 fits in the loop-shaped architecture (Fig. 1). It is worth noting that this method is slightly different from another proposal (*17*), which does not fit in the circuit shown in Fig. 1 as it requires an additional outer loop (*28*).

**Acknowledgments**

The authors acknowledge Akira Furusawa for providing space for the experiment. The authors thank Takahiro Mitani for the careful proofreading of the manuscript. The authors acknowledge Tomonori Toyoda at the Equipment Development Center of the Institute for Molecular Science for support on making electric devices through the Nanotechnology Platform Program.

**Funding:** This work was partly supported by JST PRESTO (JPMJPR1764), MEXT Leading Initiative for Excellent Young Researchers, Katsu Research Encouragement Award of the University of Tokyo, Toray Science Foundation (19-6006), Research Foundation for Opto-Science and Technology, the Kayamori Foundation of Informational Science Advancement, and the Nanotechnology Platform Program (Molecule and Material Synthesis) of MEXT, Japan. K.T. acknowledges financial support from ALPS.

**Author contributions:** Y.E. and K.Y. contributed to the adjustment, tuning, and debugging of the experimental setup, the data acquisition and analysis, and the model calculation. Y.M. contributed to the development of the optical and electrical part of the feedforward system. K.T. contributed to the development of the electrical part of the feedforward system. S.T. conceived, planned, and designed the whole system, supervising the project. All authors contributed to writing the manuscript.

**Competing interests:** The authors declare that they have no competing interests.

**Data and materials availability:** All data needed to evaluate the conclusions in the paper are present in the paper and/or the Supplementary Materials.

**Figures and Tables**

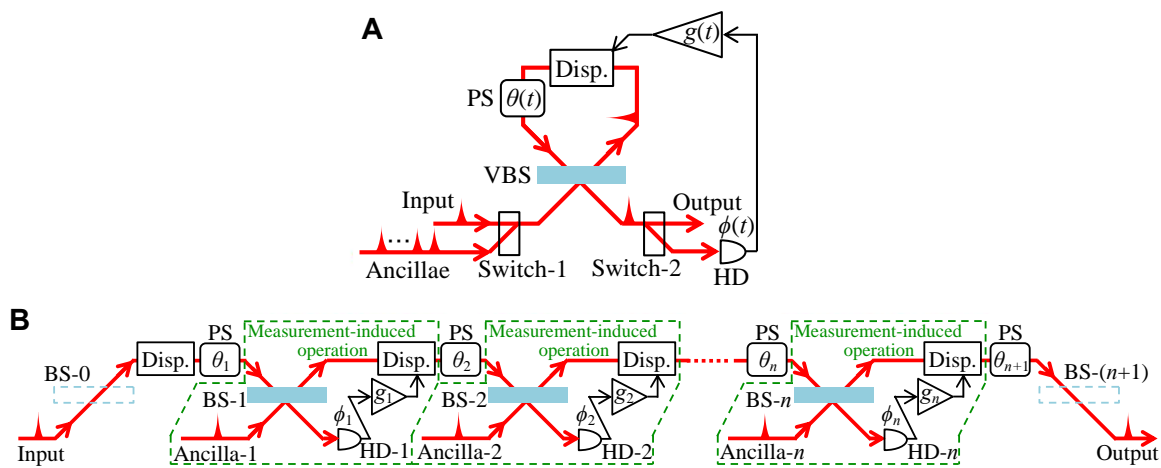

**Fig. 1. Conceptual schematic of the loop-based circuit.** (**A**) Physical configuration of the processor. The parameter of each optical and electrical component is dynamically changed as a single wave packet passes through it. VBS, variable beam splitter; PS, phase shift operation; HD, homodyne detector; Disp., displacement operation. (**B**) Equivalent circuit model of the processor. Single time-dependent components are mapped to multiple time-independent components. Temporal modes of the ancillary states are mapped to spatial modes. The amount of displacement operation can actually be dependent on all the past measurement results though this figure implies that it is determined only by the last measurement result. The measurement basis can also be dependent on the past results.

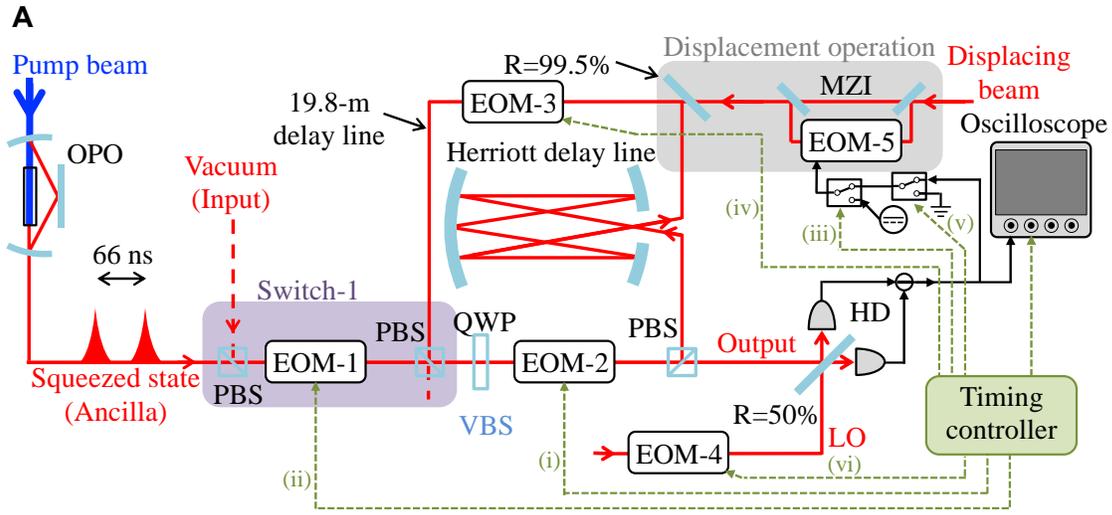

**Fig. 2. Schematic of the experiment.** (**A**) Optical and electrical setup. Refer to Materials and Methods section for the details. EOM, electro-optic modulator; LO, local oscillator; MZI, Mach–Zehnder interferometer; OPO, optical parametric oscillator; PBS, polarizing beam splitter; QWP, quarter-wave plate. (**B**) Typical timing chart of each component. This plot represents the case of a QPG with $\kappa = 0.75$ on the X-coherent state (Table 1). The gate sequence starts at around 0 ns by taking the input pulse in the loop part with the reflectivity of the VBS set to zero. The time axis is presented in the laboratory frame, which leads to apparent time deviations in particular for (v) Feedforward.

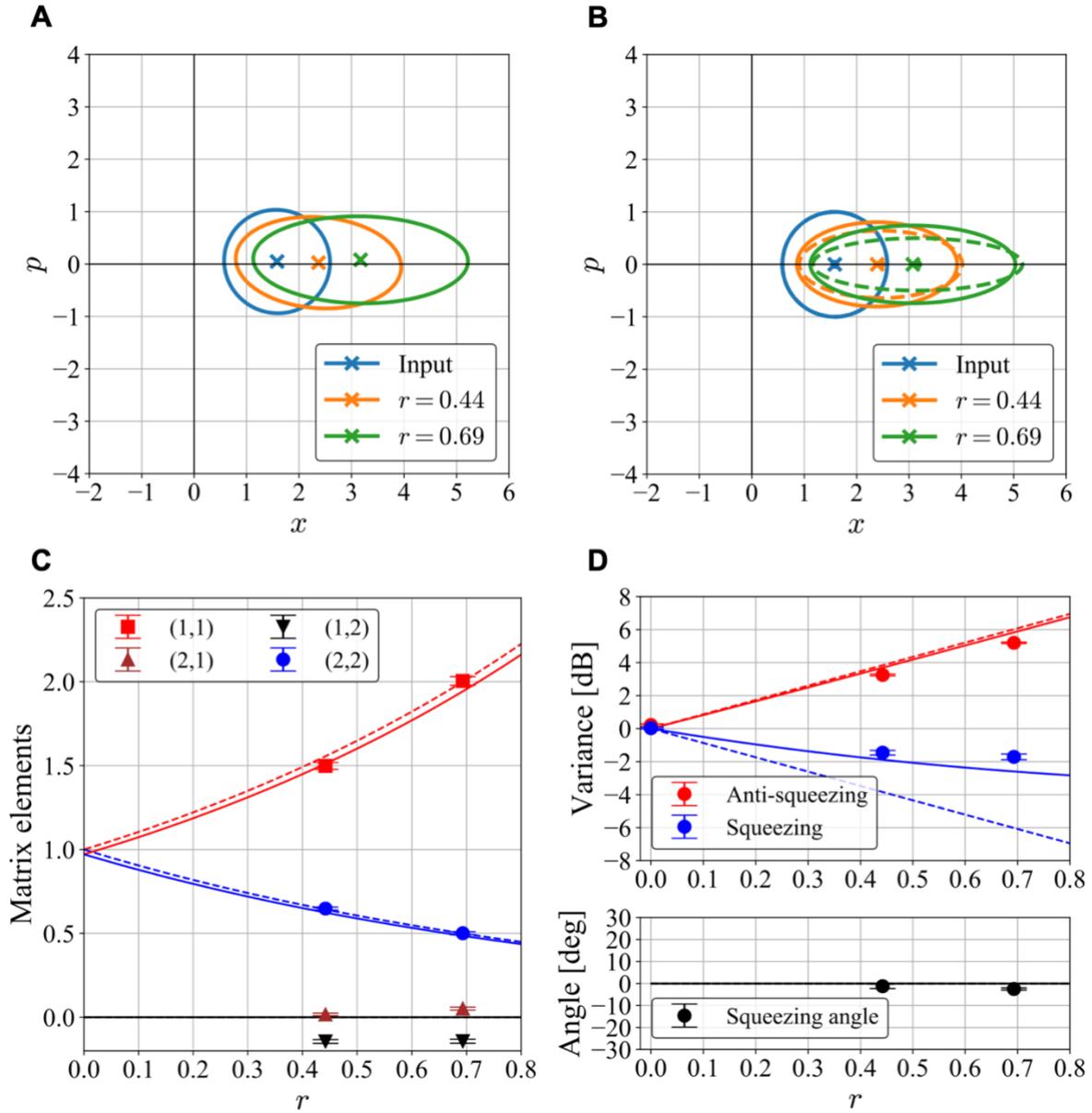

**Fig. 3. Output state characterization of the single-step squeezing gates.** (**A**) Representative Wigner functions of input and output states of the squeezing gates of the experiment. An x-shaped symbol denotes the center of the distribution, while a solid line is a contour of the relative height of $1/\sqrt{e}$ with respect to the center height representing the standard deviation. The convention $\hbar = 2$ is adopted here. (**B**) Theoretical predictions for the corresponding Wigner functions. A set of an x-shaped symbol and a solid line is derived from a realistic model with optical loss and finite squeezing, while another set of a circular dot and a dashed line is from an ideal model with no optical loss and infinite squeezing. (**C**) Matrix elements of a squeezing gate as a function of a target squeezing parameter $r$. (**D**) Parameters of a squeezing ellipse as a function of a target squeezing parameter. The variances of squeezing and anti-squeezing are shown in the upper figure while the squeezing angles are in the lower one.

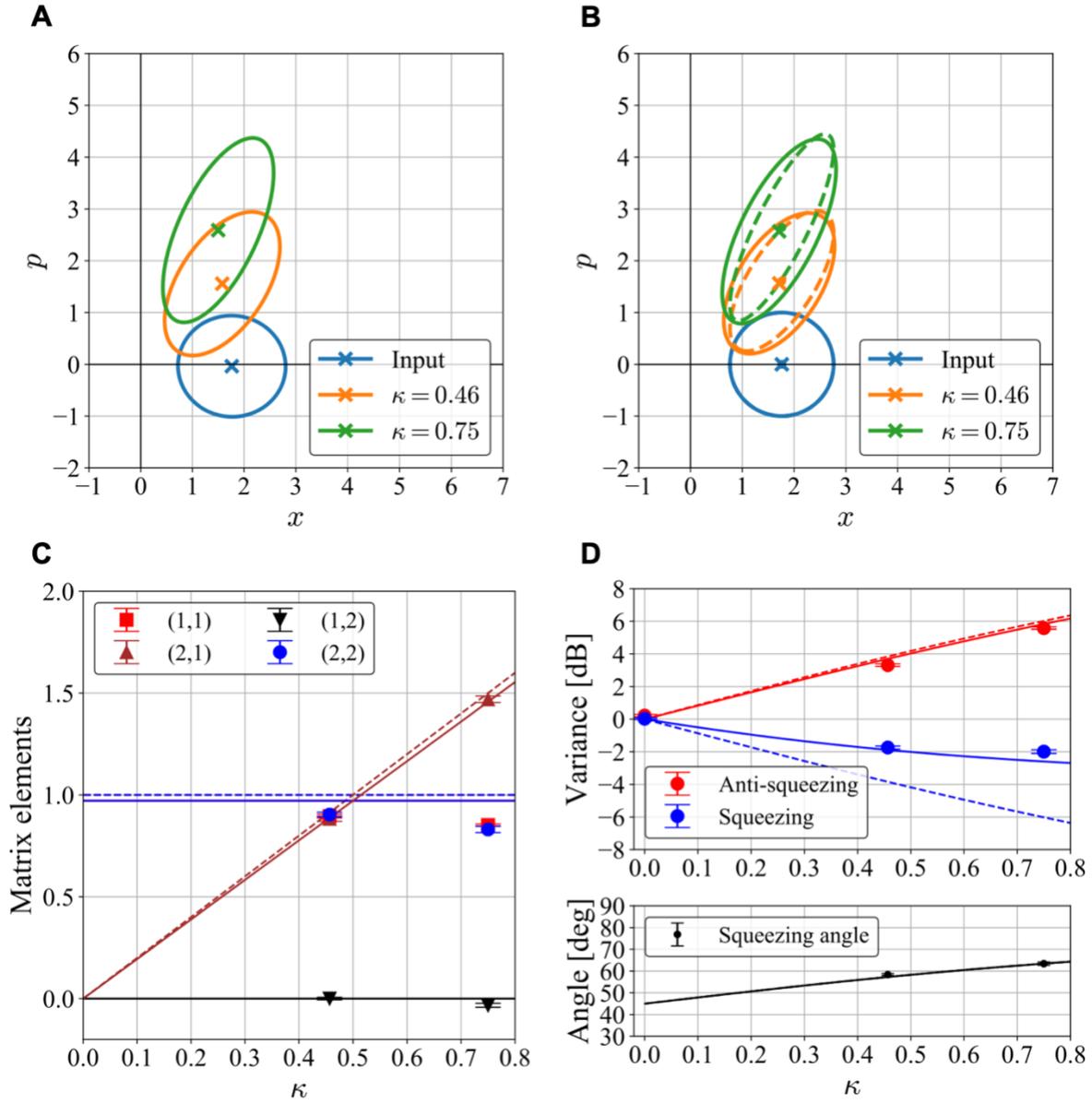

**Fig. 4. Output state characterization of the single-step quadratic phase gates.** The results are shown with the same rule as Fig. 3 except that the latter two plots are shown as functions of $\kappa$ instead of $r$.

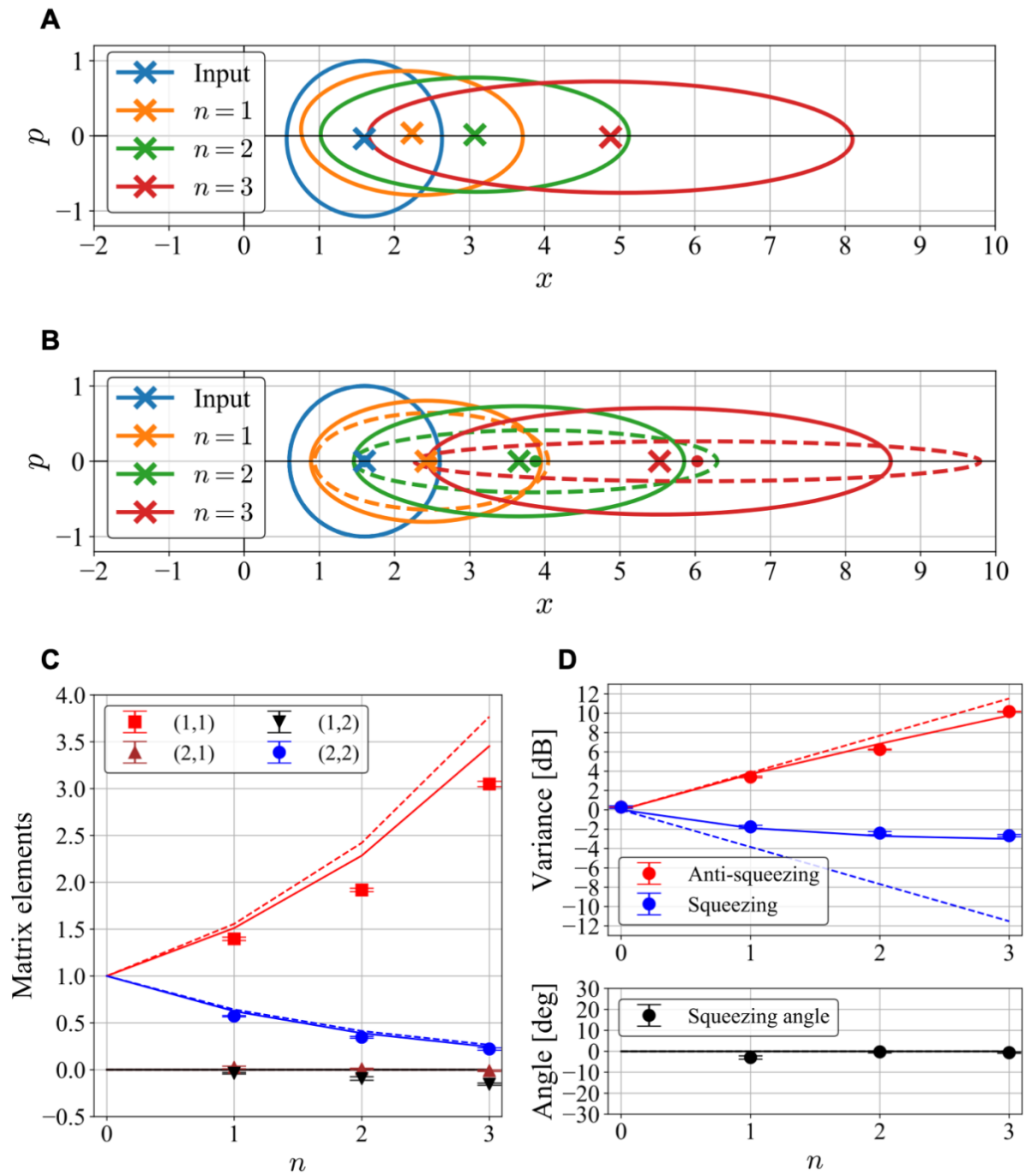

**Fig. 5. Output state characterization of the multi-step squeezing gates.** The results are shown with the same rule as Fig. 3 except that the latter two plots are shown as functions of $n$, the number of repeats.

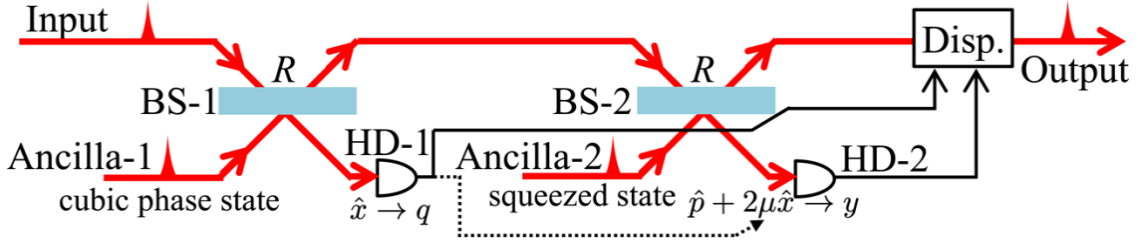

**Fig. 6. Conceptual schematic of the implementation method of a cubic phase gate.**
The input state to this circuit is processed in a measurement-induced method using a cubic phase state and a squeezed state as ancillae. The homodyne angle of HD-2 is adaptively controlled according to the outcome of HD-1. The displacement operation is determined by both of the results from HD-1 and HD-2.

**Table 1. Parameter list of the optical components for the demonstration of Gaussian gates.** The first column identifies the target gate. $R$(BS-1) is the reflectivity of BS-1 in the virtual diagram shown in Fig. 1B. $\theta_1$ and $\theta_2$ are the amount of phase shifts, which can also be found in Fig. 1B. $g_1$ is the ratio of the amount of displacement with respect to the value of the measured quadrature. The homodyne angle $\phi_1$ is not listed here since it is 0 deg in all cases.

| $r$ or $\kappa$ | $R$(BS-1) (%) | $\theta_1$ (deg) | $\theta_2$ (deg) | $g$ (dB) |
|---|---|---|---|---|
| $r = 0.44$ | 41 | 0 | 0 | 1.6 |
| $r = 0.69$ | 25 | 0 | 0 | 4.8 |
| $\kappa = 0.46$ | 41 | $-32.7$ | 57.3 | 1.6 |
| $\kappa = 0.75$ | 25 | $-26.6$ | 63.4 | 4.8 |

# Supplementary Materials for

## Programmable and sequential Gaussian gates in a loop-based single-mode photonic quantum processor

Yutaro Enomoto, Kazuma Yonezu, Yosuke Mitsuhashi, Kan Takase, Shuntaro Takeda*

*Corresponding author. Email: takeda@ap.t.u-tokyo.ac.jp

**Supplementary Text**

<u>Electro-optic modulators</u>
We use three different types of electro-optic modulators (EOMs). EOM-1 and EOM-2 (Fig. 2A) are bulk polarization modulators, RTP-X-4-20-AR650-1000-HV from Laysop Ltd. For driving them, we use high voltage drivers, custom-made versions of PCD-bpp, from Bergmann Messgeräte Entwicklung KG. They rotate the polarization angle of the incoming beam depending on the applied voltage. EOM-3 is a bulk phase modulator, RTP-PM-X-4-20-AR650-1000-HV from Laysop Ltd. We use the same driver for driving it. It shifts the optical phase of the incoming beam. EOM-4 and EOM-5 are fiber-coupled phase modulators, NIR-MPX800-LN-05-P-P-FA from iXblue Photonics, which do not require a special driver.

EOM-2 is a variable polarization rotator and sandwiched with two polarization beam splitters (PBSs), constituting the variable beam splitter (VBS) (*29*). While a set of EOM-2 and two PBSs alone serves as a VBS, we also place a quarter-wave plate between two PBSs. This is intended to set the initial reflectivity of the VBS to 50 % with no voltage applied to EOM-2, making it easy for us to control the optical path length of the loop part; the optical path length can be sensed by a modulation and demodulation technique since the loop part can be seen as an optical ring cavity whose input coupler has a reflectivity of 50 %. We choose 50 % because it is realized simply by inserting a quarter-wave plate though any reflectivity but 0 % or 100% is acceptable for the control. The path length is corrected by a mirror equipped with a piezoelectric actuator.

With the current driver for EOM-2, two independent values of the voltage, $V_1$ and $V_2$, can be programmed so that four different values of $0, V_1, V_2$ and $V_1 + V_2$ are applied to EOM-2 (*29*). One degree of freedom, namely $V_1$, is used to make the reflectivity of the VBS be 0 % when the loop imports the input state or exports the output state. Thus the number of the remaining degree of freedom is one. This is the reason why we choose to apply the same squeezing gates over the multiple steps for the demonstration. However, this is not a fundamental limitation. As discussed in the reference (*29*), it can be overcome by developing a more sophisticated driver or cascading the same EOMs.

EOM-4 is used to vary the homodyne angle by changing the applied voltage when LO shift is ON (Fig. 2). This function is necessary for the quantum tomography of the output state of the processor. The applied voltage ranges between − 4.59 V and + 0.86 V corresponding to – 135 degrees and + 30 degrees, respectively. For each homodyne angle, we repeat the measurement sequence 1000 times, as a representative one is shown in Fig. 2B.

Displacing beam

The displacing beam (Fig. 2A) has two functionalities. One is the optical feedforward to finalize the quantum operations. The other is the preparation of the input coherent state. In our demonstration, instead of externally injecting coherent states from Switch-1, we internally prepare the coherent states in the loop part, where the displacing beam displaces the vacuum state taken in from Switch-1. When the displacing beam is used for the feedforward, the electrical switches are configured so that the outcome of the homodyne detector is sent to EOM-5 ((v) in Figs. 2A and 2B). On the other hand, when it is used for the input state preparation, the constant voltage source is connected to EOM-5 ((iii) in Figs. 2A and 2B). Since the mode function (Eq. 3) is anti-symmetric with respect to its center, constant displacement over the whole time bin results in no net displacement. Thus, we enable the displacement only for 20 ns within the first half of the time bin (Fig. 2B (iii)) to produce coherent states from the vacuum state.

The Mach–Zehnder interferometer (MZI) in the path of the displacing beam removes the carrier field of the displacing beam by a destructive interference. The carrier field produces an offset in the outcome of the homodyne detector. Ideally, the carrier field will not affect the experimental results thanks to the insensitivity of the mode function (Eq. 3) to the overall offset. However, it does affect to some extent in reality since the amplitude of the carrier field is too large without the MZI. Moreover, due to the dynamical change of the reflectivity of the VBS and thus the offset level originating from the carrier field, the offset cannot be removed simply by applying a high-pass filter to the outcome of the homodyne detector. This is the reason why we implement the MZI to reduce the carrier field to mitigate its undesired effect.

Preliminary measurements for time synchronization and parameter calibration

The synchronization of each component is adjusted by preliminary measurements. The timing of each switching is adjusted in the following way. The classical light fields are introduced into the optical circuit, and the fields are measured by detectors at the output port. A pulse-like switching signal to each component produces a pulse in the time series of the outcome of the detectors. By measuring the timing of the pulse in the time series with respect to the trigger pulse, we can infer when to send switching signals to each component. As the timing controller can create pulse signals with controlled delays, it sends the switching signals at the inferred timings with one nanosecond precision. The electrical delay between the homodyne detector and EOM-5 is adjusted by the length of the coaxial cable transmitting the feedforward signal.

The reflectivity of the VBS is calibrated in advance in a similar way with the classical light field from the injection port for the ancillary states. By measuring the optical power at the output port with the voltage applied to EOM-2 varied, the correspondence between the reflectivity and the voltage level is obtained. The amounts of the phase shift by EOM-3 and -4 are also calibrated in the similar way. By sinusoidal fittings of the outcome of the detector, the correspondence between the phase shift and the voltage level applied to EOM-3 or -4 is obtained.

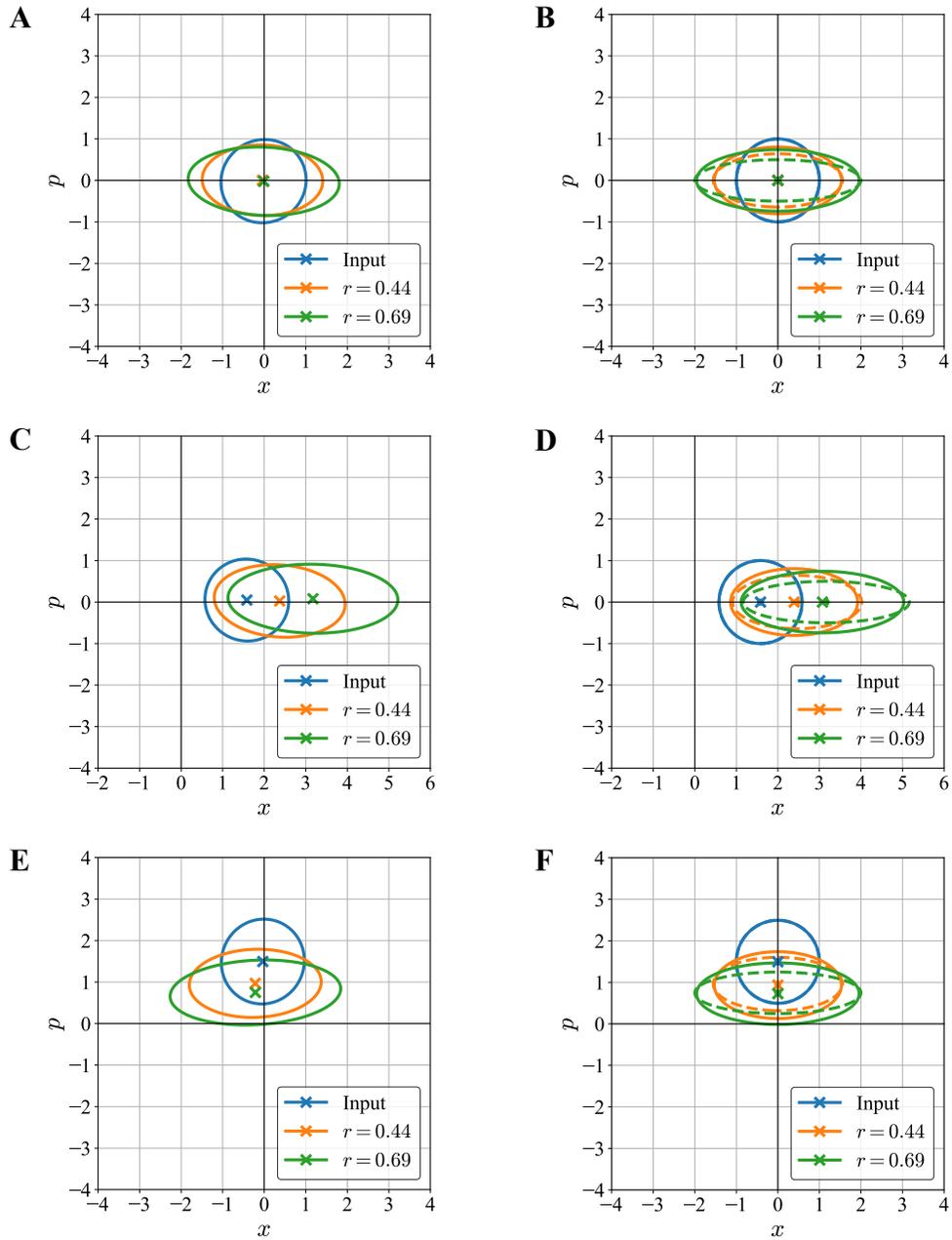

**Fig. S1.**

Wigner functions of input and output states of the single-step squeezing gates ($r = 0.44$ and $0.69$) shown with the same rule as Fig. 3. (**A**), (**C**), and (**E**) show the experimental results while (**B**), (**D**), and (**F**) show the theoretical predictions. (**A**) and (**B**) are for the vacuum input state, (**C**) and (**D**) are for the X-coherent input state, and (**E**) and (**F**) are for the P-coherent input state.

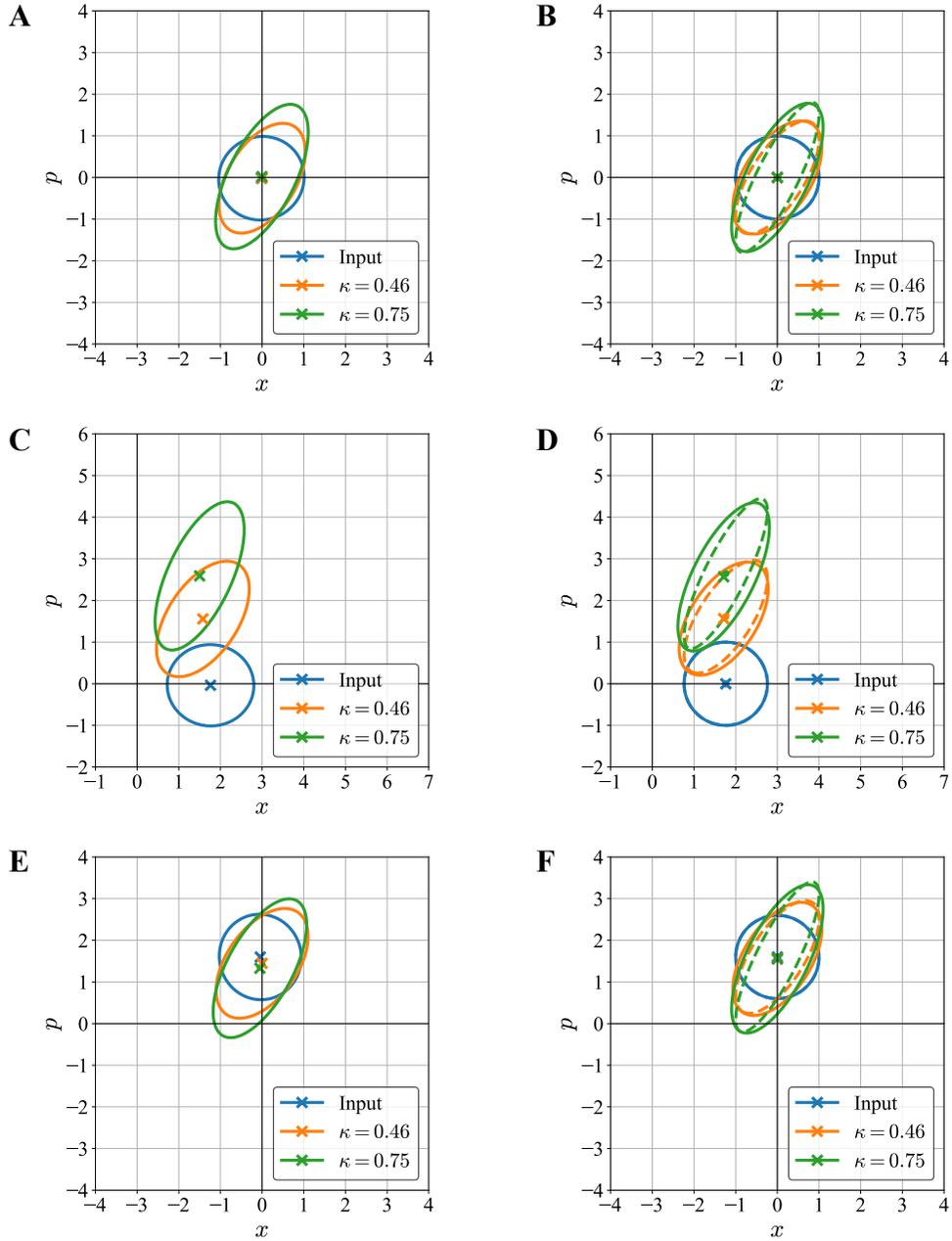

**Fig. S2.**

Wigner functions of input and output states of the single-step quadratic phase gates ($\kappa = 0.46$ and $0.75$) shown with the same rule as Fig. 4. (**A**), (**C**), and (**E**) show the experimental results while (**B**), (**D**), and (**F**) show the theoretical predictions. (**A**) and (**B**) are for the vacuum input state, (**C**) and (**D**) are for the X-coherent input state, and (**E**) and (**F**) are for the P-coherent input state.

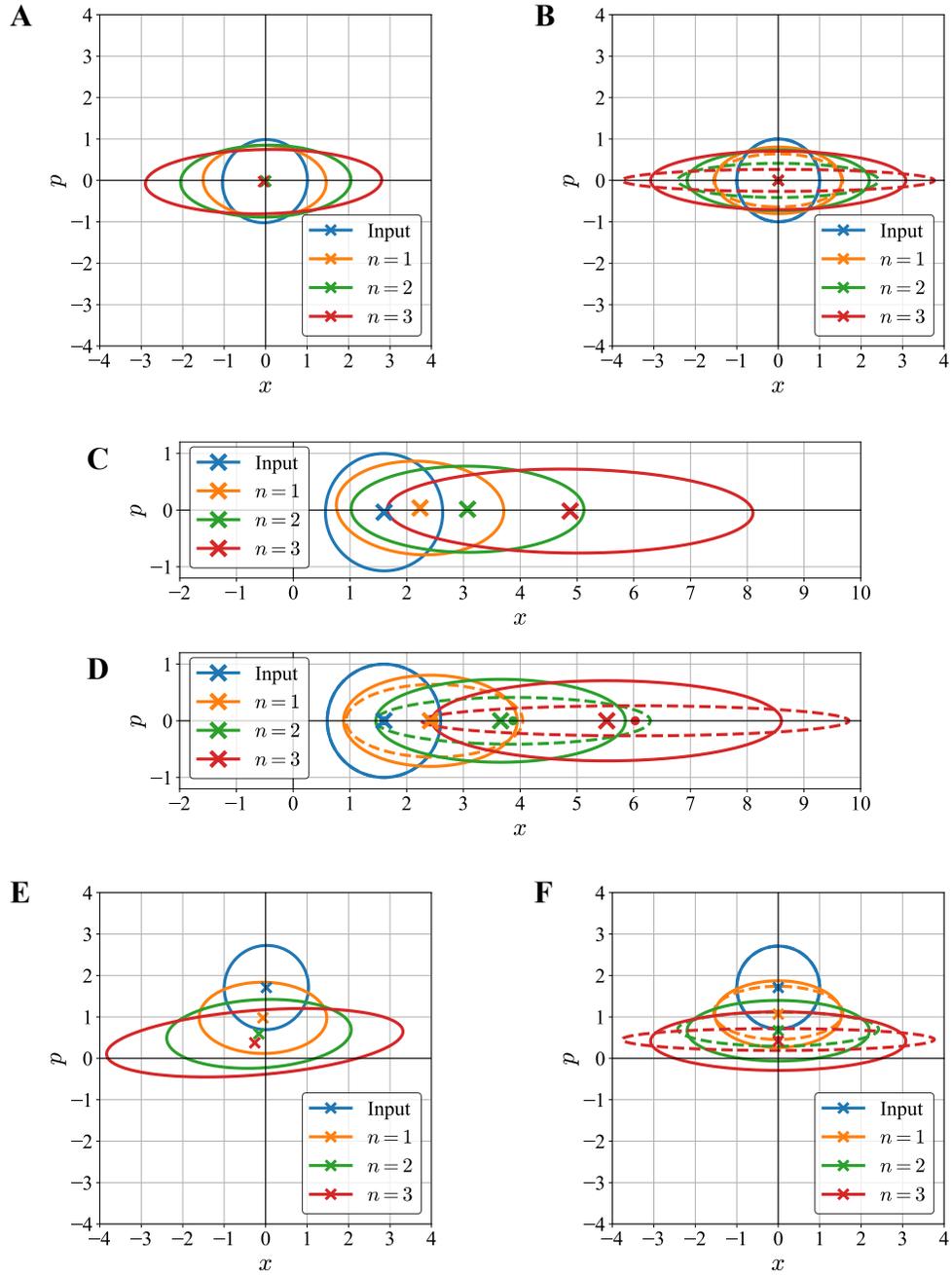

**Fig. S3.**

Wigner functions of input and output states of the multi-step squeezing gates shown with the same rule as Fig. 5. A squeezing gate with $r = 0.44$ is applied $n$ times ($n = 1, 2,$ and $3$). (**A**), (**C**), and (**E**) show the experimental results while (**B**), (**D**), and (**F**) show the theoretical predictions. (**A**) and (**B**) are for the vacuum input state, (**C**) and (**D**) are for the X-coherent input state, and (**E**) and (**F**) are for the P-coherent input state.